\documentclass[12pt]{iopart}

\usepackage{amsfonts,amssymb}
\usepackage{graphics,graphicx,epsfig}
\usepackage{amsthm}
\usepackage{color}

\newcommand{\ket}[1]{ | #1 \rangle}

\renewcommand{\epsilon}{\varepsilon} \newcommand{\avg}[1]{\langle #1\rangle}

\begin{document}

\title{Highly noise resistant multiqubit quantum correlations}

\author{Wies{\l}aw Laskowski}
\address{Institute of Theoretical Physics and Astrophysics, University of Gda\'nsk, PL-80-952
Gda\'nsk, Poland}

\author{Tam\'as V\'ertesi}
\address{Institute for Nuclear Research, Hungarian Academy of Sciences, H-4001 Debrecen, P.O. Box
51, Hungary}
\address{D\'epartement de Physique Th\'eorique, Universit\'e de Gen\`eve, 1211 Gen\`eve, Switzerland}

\author{Marcin Wie\'sniak}
\address{Institute of Theoretical Physics and Astrophysics, University of Gda\'nsk, PL-80-952
Gda\'nsk, Poland}

\submitto{\JPA}

\begin{abstract}
We analyze robustness of correlations of the $N$-qubit GHZ and Dicke states against white noise admixture. For
sufficiently large $N$, the Dicke states (for any number of excitations) lead to more robust violation of local realism than the GHZ states (e.g. for $N>8$ for the W state). We also identify states that are the most resistant to white noise. Surprisingly, it turns out that these states are the GHZ states augmented with fully product states.
Based on our numerical analysis conducted up to $N=8$, and an analytical formula derived for any $N$ parties, we conjecture that the three-qubit GHZ state augmented with a product of $(N-3)$ pure qubits is the most robust against white noise admixture among any $N$-qubit state. As a by-product, we derive a single Bell inequality and show that it is violated by all pure entangled states of a given number of parties. This gives an alternative proof of Gisin's theorem.
\end{abstract}


\maketitle

\section{Introduction}

Local realism is a model introduced by Einstein, Podolsky, and Rosen \cite{EPR} in an attempt to reach agreement between
predictions of quantum mechanics with our classical intuition. However, in a simple argument Bell has demonstrated \cite{Bell}
that this attempt must necessarily fail. Not only has the contradiction between the assumptions of local realism and quantum
mechanics deep fundamental meaning, it also lies at the very heart of more efficient solutions to certain specific communication
tasks \cite{bellreview}.

It is then natural to confront these two contradicting theories in an experiment. For the Bell theorem, many systems have been
studied, but quantum interferometry is the most common field of implementation. However, any experiment will suffer from some
imperfections, which could be both of a technical nature (to be removed by more effort of the experimentalist) and intrinsically
irremovable. They degrade the quality of quantum correlations, making them more similar to those reproducible by local operations and classical communication. This fact is reflected in the theoretical description by mixing the desired state with a certain noise. The simplest noise model is the white noise. It has a uniform structure and degrades all correlations of the state by a constant factor, $v$, which corresponds to Michelson's interference visibility.

Robustness of Bell inequality violation against white noise is often
considered as a benchmark of the strength of nonclassicality of states. In this contribution we conduct a numerical investigation of multiqubit states that are the most robust against this imperfection, in the framework of two measurements per side. We find that the most noise is tolerated in protocols similar to those described in Ref.~\cite{SEN}, where a group of observers simply projects on fixed local states. More surprisingly, the most robust states are composed of the GHZ and pure one-qubit states.

Lately, considerable research effort has been dedicated to the study of noise robustness of multipartite entangled states (see, e.g., \cite{NLclasses,decoherenceNL,dickenoise,decoherenceENT, szumy}). Specifically, in Ref.~\cite{NLclasses}, families of Bell inequalities have been constructed for discriminating different types of multipartite entanglement (e.g. GHZ~\cite{GHZ} and W-type~\cite{Dicke} of entanglement). The present study can be considered as part of this research program.

\section{Tools}

In order to analyze noise properties of $N$-qubit quantum states one can define a ``critical visibility'' \cite{KASZLIKOWSKI} as the value of the parameter $v$, non-negative, less than or equal to 1, for which, under a fixed set of conditions
(number of observers, number of settings, and kinds of observables), a mixed state defined by
\begin{equation}
\label{whitenoise}
\rho(v) = v \rho + \frac{1-v}{2^N} I.
\end{equation}
loses the nonclassical properties of the original state $\rho$. Actually, Eq.~(\ref{whitenoise}) describes a situation when the initial state $\rho$ goes through a global depolarizing channel. One can also consider a situation where the white noise affects each qubit independently corresponding to a local depolarization. Entanglement and nonlocal properties of such channels have been recently studied in Refs.~\cite{decoherenceENT} and \cite{decoherenceNL}, respectively.

Our task is to find, for a given state $\rho$, the critical value $v_{crit}$ such that if $v > v_{crit}$, there does not exist {\em any} local realistic model describing quantum correlations of experimental events.

This is done using a numerical method based on linear programming.
The numerical procedure was first used in \cite{KASZLIKOWSKI} to show that violation of local realism is stronger (more white noise resistant) for two qu$d$its ($d>2$) than for two qubits. Later the method was also used for the analysis of violations of local realism by different kinds of multiqubit quantum states,  which are often discussed in the context of quantum information \cite{SR1}, two qutrits in all possible pure entangled states \cite{SR2} and some families of two qutrits states \cite{SR4}. The method is described and discussed in details in \cite{SR1, SR3}.

In our numerical analysis $N$ spatially separated observers perform measurements of two alternative dichotomic observables.
In this case the local realistic models are equivalent to the existence of a joint probability distribution $p_{\rm LR}(a_1^{(1)}, a_2^{(1)}, \dots, a_1^{(N)}, a_2^{(N)})$, where $a_j^{(i)} =\pm 1$ denotes the result of the measurement of the $j$th observable performed by the $i$th observer.

A set of $2^{2N}$ quantum probabilities $p_{\rm QM}(r_1, \dots, r_N | A^{(1)}_{i_1}, \dots, A^{(N)}_{i_N})$ of obtaining the result $r_j$ by $j$th observer, measuring
the observable $A_{i_j}^{(j)}$ ($i_j = 1,2$; $j=1,\dots,N$), corresponds to a single point in a $2^{2N}$-dimensional space. Those sets which can be described using local realistic models form a convex geometric figure called a Bell-Pitovsky polytope \cite{PITOVSKY} in this $2^{2N}$-dimensional space. The facets of the polytope are equivalent to tight Bell inequalities. For $v = v_{crit}$ quantum predictions for the probabilities $p_{\rm QM}$ are the marginal sums of the joint probability distribution $p_{\rm LR}$ and form the coordinates of  a geometric point which lies in a facet of the polytope (i.e., they saturate a Bell inequality). It has to be noted, that our numerical method does not need any knowledge at all of the forms of Bell inequalities. However, the results obtained by this method are equivalent to the analysis of the full set of (probabilistic) Bell inequalities formulated for a given problem. For some cases, the Bell inequalities that are violated by the states are specified using the method of \cite{ELIOT}. This method enables us to extract a facet-defining Bell inequality from the correlation point provided by the linear programming method for a given scenario.

\section{Properties of the noisy multipartite GHZ and Dicke states}

At the beginning we analyze how correlations of two prominent families of states are resistant to white noise admixture. The
states which we consider are: the $N$-qubit GHZ state \cite{GHZ}: \begin{equation} \ket{{\rm GHZ}_N} =
\frac{1}{\sqrt{2}}(\ket{0\dots 0} + \ket{1\dots 1}) \end{equation} and the $N$-qubit Dicke states (with $e$ excitations)~\cite{Dicke}:
\begin{equation}
\ket{D^e_N} = \frac{1}{\sqrt{{N \choose e}}} \sum_{\pi} \ket{\pi(\underbrace{0 \dots 0}_{N-e}\underbrace{1 \dots 1}_{e})},
\label{DEN}
\end{equation}
where $\pi$ denotes a permutation of $e$ ones and $(N-e)$ zeros in the ket, and ${N \choose e}$ gives
a number of such permutations. The special case of $e=1$ corresponds to the $N$-qubit W states.

\subsection{Numerical method}

We applied the numerical method and found critical visibilities for the GHZ and W ($D^1_N$) states up to 11 qubits in experiments with two alternative measurement settings per side.  The results are presented in Table~\ref{ghzw}.

\begin{table}
\caption{\label{ghzw} The critical visibilities for the GHZ and W states for $N\leq 11$ and two measurement settings per observer. If $v > v_{crit}$, there does not exist any local realistic model describing quantum probabilities of experimental events. The W states lead to lower critical visibility than the GHZ states for $N > 8$, whereas the projective method \cite{SEN} shows this fact for $N \geq 11$.}
\footnotesize \rm
\begin{tabular*}{\textwidth}{@{}l*{15}{@{\extracolsep{0pt plus12pt}}l}} \br
$N$ & $v_{crit}^{GHZ}$ & $v_{crit}^{W}$ & projective  method \cite{SEN} \\ \mr
3&   0.5000& 0.6442& 0.6442\\
4&   0.3536& 0.5294& 0.5469\\
5&   0.2500& 0.4018& 0.4300\\
6&   0.1768& 0.2774& 0.3116\\
7&   0.1250& 0.1736& 0.2089\\
8&   0.0884& 0.1034& 0.1311\\
9&   0.0625& {\bf 0.0578} & 0.0782\\
10&  0.0442& 0.0313& 0.0450\\
11&  0.0313& 0.01687 & {\bf 0.0252} \\
\br
\end{tabular*}
\end{table}

Analyzing the critical values one can see that the W state is more resistant to white noise than the GHZ state already for $N>8$. This result is stronger than the one obtained by means of the projective method presented in \cite{SEN}.  In this method, applied for the $N$-qubit W state mixed with white noise, if measurements at (N-2) parties are made in the computational basis and all yield the ``+1'' result (associated with $\ket{0}$) the remaining pair of observers is left with a mixture of a 2-qubit Bell state ($\ket{\psi^+}$) with a  reduced amount of white noise. 
The critical visibilities for the W state obtained using that method are presented in the last column of Tab. \ref{ghzw} and they are higher than the corresponding critical visibilities obtained by the numerical method (the second column) by about $3\%$ (for $N = 4$) to $50\%$ (for $N = 11$). The projective method predicts that the W state is more resistant to white noise than the GHZ state only for $N \geq 11$.

Note that due to the significant improvement of the accuracy of the numerical method, the critical visibilities for $N$-qubit W state ($4 \leq N \leq 7$) are lower (in the second decimal place) than those calculated previously with a similar numerical method and presented in \cite{SEN, SR1}.

\subsection{A family of Bell inequalities}
\label{family}

It is well known that the GHZ states exhibit maximal violation of the Bell inequalities \cite{Mermin, WWWZB} for experiments with two alternative
measurements setting per party.  A noisy $N$-qubit GHZ state leads to the critical visibility
\begin{equation}
v_{crit}^{GHZ} = \frac{1}{2^{(N-1)/2}}.
\label{vGHZ}
\end{equation}
Actually, the set of inequalities in \cite{WWWZB} comprises all tight two-setting full-correlation-type Bell inequalities, hence the above threshold (\ref{vGHZ}) for the visibilities defines the lowest one among all such correlations. However, we cannot definitely exclude that for $N>11$ there are no probabilistic inequalities that lead to the critical visibility lower than (\ref{vGHZ}).

For the Dicke states $D_N^e$ ($N \geq 3$) we introduce a new Bell inequality, which is a facet of the Bell-Pitovsky polytope in the case of $D_3^1$ state. The inequality is obtained from results of the numerical method using a procedure described in \cite{ELIOT} and then extended to $N$ parties.  The inequality has the following iterative form
\begin{equation}
\langle C_N \rangle = \langle (1-A^{(N)}_1) C_{N-1} + 2 A^{(N)}_1 \rangle \leq 2,
\label{ineq_dicke}
\end{equation}
where $C_2  =
A^{(1)}_1 A^{(2)}_1+A^{(1)}_1 A^{(2)}_2+A^{(1)}_2 A^{(2)}_1-A^{(1)}_2 A^{(2)}_2$ is the CHSH expression \cite{CHSH} and
$A^{(i)}_j$ denotes a dichotomic observable measured by $i$th observer when he/she chooses $j$th measurement setting. Note that
($N-2$) observers perform only a single measurement $A^{(k)}_1~(k=3,\dots,N)$. The Bell expression (\ref{ineq_dicke}) can be
explicitly written as
\begin{eqnarray}
C_N &=& C_2 + (1 - C_2) \left(\sum_{i=3}^N A^{(i)}_1 - \sum_{3 \leq i < j \leq N}
A^{(i)}_1 A^{(j)}_1 \nonumber  \right.\\
    &+&  \sum_{3 \leq i < j < k \leq N} A^{(i)}_1 A^{(j)}_1 A^{(k)}_1 - \cdots
		\label{ineq_dicke1} + \left. (-1)^{N+1} A^{(3)}_1 ... A^{(N)}_1 \right).\nonumber
\end{eqnarray}
We also mention that this inequality reduces to the one appeared in \cite{BV} for $N=3$.

In order to find the quantum value of (\ref{ineq_dicke1}) let us choose measurement settings for the last ($N-2$) observers
$A^{(i)}_1 =-\sigma_z$ $(i=3,\dots, N)$. Note that due to the permutational symmetry of the Dicke states the correlations are the
same for any particular set of subsystems. The quantum value of the inequality (\ref{ineq_dicke1}) can read
\begin{eqnarray}
\avg{C_N}_{D_N^e} &=& \avg{C_2}_{D_N^e} + (1 - \avg{C_2}_{D_N^e}) \sum_{k=1}^N (-1)^{2k+1} {N \choose k}
T_{\underbrace{z...z}_{k}\underbrace{0...0}_{N-k}},
\end{eqnarray}
where the expectation value of the CHSH operator is equal to
$\avg{C_2}_{D_N^e} = 2\sqrt{2} \cdot v \cdot \frac{2}{{N \choose e}}$ and \begin{eqnarray} T_{\underbrace{z...z}_{k}0...0} =
\frac{v}{{N \choose e}} \sum_{j=0}^{\left\lceil e/2\right\rceil} \Big[{n-k\choose e-2j} {k\choose 2j}\nonumber  -{n-k\choose
e-(2j+1)}{k\choose 2j+1}  \Big].
\end{eqnarray}
After calculations we get
\begin{equation} \avg{C_N}_{D_N^e} = v
\left(2+\frac{2^{N} (\sqrt{2}-1)}{{N \choose e}}\right). \end{equation} Therefore, the critical visibility is equal to
\begin{equation} v_{crit}^{_{D_N^e}} = \left(1+\frac{2^{N-1} (\sqrt{2}-1)}{{N \choose e}}\right)^{-1}. \label{vDicke}
\end{equation}
Comparing Eqs.~(\ref{vGHZ}) and (\ref{vDicke}) we can conclude that for any type of the Dicke state $D_N^e$ (for an arbitrary number of excitations $e$), there is a critical number of particles $N^e_{crit}$ such that $v_{crit}$ for ${D_{N_{crit}}^e}$ is lower than critical visibility for the $N$-qubit GHZ state. For example: for $e=1$, $N^1_{crit}=11$; for $e=2$, $N^2_{crit}=19$;  for
$e=5$, $N^3_{crit}=44$; for $e=10$, $N^4_{crit}=88$. It means that for $N>N^e_{crit}$ the Dicke states $D_N^e$ become more robust against white noise admixture than the GHZ states in terms of violation of local realism. For $e=1$, Eq. (\ref{vDicke}) recovers the results obtained with the projective method \cite{SEN} (see the last column of Tab. \ref{ghzw}). These results are in good agreement with the recent ones~\cite{dickenoise} where the most robust Dicke states subject to losses are found to correspond to only few excitations $e$.

It is worth mentioning that after a simple modification of the inequality (\ref{ineq_dicke}), one can obtain a three-setting Bell inequality which is violated by {\em any} pure entangled state. See~\ref{app:ineq} for details.

\section{Quantum states with the highest resistance to white noise}

In this section we identify states that lead to the highest robustness against white noise admixture for $N$ observer scenario ( for $N \leq 6$) and give a conjecture for any arbitrary $N$.

\subsection{The optimal case: three qubit GHZ state augmented with product states}

We also apply the linear programming method to identify states that lead to the lowest critical visibility for $2 \leq N \leq 6$. This was possible after including an optimization over {\em all pure states} in the numerical method. The optimal states are: for $N=2$, $\ket{{\rm GHZ}_2}$; for $N=3$, $\ket{{\rm GHZ}_3}$; for $N=4$,  $\ket{{\rm GHZ}_3}\ket{0}$; for $N=5$, $\ket{{\rm GHZ}_3}\ket{00}$; and for $N=6$,
$\ket{{\rm GHZ}_3}\ket{000}$. These states are the most robust states against white noise admixture for a given number of qubits. This means that there are no other states violating any Bell inequality in a presence of more noise.


Additionally we analyze noise resistance of the states composed of the $r$-qubit GHZ state and $(N-r)$-product state
\begin{equation}
|\psi_{r}^N \rangle = |{\rm GHZ}_{r} \rangle|\underbrace{0 \dots 0}_{N-r} \rangle.
\label{state_prod}
\end{equation}
The results for two measurement settings per party are presented in Tab. \ref{product}.

\begin{table}
\caption{\label{product} Comparison of the critical visibilities for the augmented GHZ states of the form (\ref{state_prod}) with corresponding values for the full $N$-qubit GHZ states for experiments with two measurement settings per observer. According to our numerical analysis, the states that appear in bold for a given number of qubits $N\le 6$ are the most resistant to white noise among all pure $N$-qubit states.}
\footnotesize\rm
\begin{tabular*}{\textwidth}{@{}l*{15}{@{\extracolsep{0pt plus12pt}}l}} \br
$N$ & State & $v_{crit}$ \\ \mr
2 & $\ket{{\rm GHZ}_2}$ & {\bf 0.707}\\ \mr
3 & $\ket{{\rm GHZ}_3}$ & {\bf 0.500}\\
& $\ket{{\rm GHZ}_2}\ket{0}$ & 0.547\\ \mr
4 & $\ket{{\rm GHZ}_4}$ &  0.354\\
& $\ket{{\rm GHZ}_3}\ket{0}$ & {\bf 0.333}\\
& $\ket{{\rm GHZ}_2}\ket{00}$ & 0.377\\ \mr
5 &$\ket{{\rm GHZ}_5}$ & 0.250 \\
&$\ket{{\rm GHZ}_4}\ket{0}$  & 0.215 \\
&$\ket{{\rm GHZ}_3}\ket{00}$ & {\bf 0.200} \\
&$\ket{{\rm GHZ}_2}\ket{000}$ & 0.232 \\ \mr
6 &$\ket{{\rm GHZ}_6}$ & 0.177 \\
&$\ket{{\rm GHZ}_5}\ket{0}$  &  0.144\\
&$\ket{{\rm GHZ}_4}\ket{00}$ & 0.121\\
  &$\ket{{\rm GHZ}_3}\ket{000}$ &  {\bf 0.111}\\
  &$\ket{{\rm GHZ}_2}\ket{0000}$ & 0.131 \\ \mr
7 &$\ket{{\rm GHZ}_7}$ &  0.125 \\
  &$\ket{{\rm GHZ}_6}\ket{0}$  & 0.098 \\
  &$\ket{{\rm GHZ}_5}\ket{00}$ & 0.078 \\
  &$\ket{{\rm GHZ}_4}\ket{000}$ & 0.064\\
  &$\ket{{\rm GHZ}_3}\ket{0000}$ & {\bf 0.059} \\
  &$\ket{{\rm GHZ}_2}\ket{00000}$ & 0.070 \\ \mr
8 &$\ket{{\rm GHZ}_5}\ket{000}$ & 0.041 \\
  &$\ket{{\rm GHZ}_4}\ket{0000}$ & 0.033 \\
  &$\ket{{\rm GHZ}_3}\ket{00000}$& {\bf 0.030} \\
  &$\ket{{\rm GHZ}_2}\ket{000000}~~~~~$& 0.036 \\
  &$\ket{{\rm GHZ}_8}$& 0.088\\
\br \end{tabular*}
\end{table}

Surprisingly, the obtained critical parameters $v_{crit}$ are very low and lower than the corresponding ones for the full $N$-qubit GHZ state. The surprise is due to the fact that the $N$-qubit GHZ state is genuinely $N$-partite entangled and maximizes many entanglement conditions and measures \cite{HORODECKI-review}.  Our result is also consistent with the observations presented in Ref. \cite{decoherenceENT}. There it was shown that the entanglement of the $N$-qubit GHZ state decays exponentially with the number of qubits in the presence of local white noise.

Notice that it is relevant that we have augmented the GHZ state with pure single-qubit states, while the white noise is always admixed. As a result, we get a state, which is partially separable, but not partially product. Every successful projection on $|0\rangle$ by one of $(N-3)$ observers, who share a non-entangled part of the state, reduces the ratio between the weights of the white noise and the state by a factor of $2$. In this way, we are able to filter out the noise from the entangled state. It is straight-forward to see that such a supplement is optimal. The filtering effect is much more prominent than in Ref. \cite{SEN}. Therein, robustness of the Bell non-classicality was studied for multiqubit W states, which do not have $|\langle\sigma_z\rangle|=1$ for any individual particle, and the decay of the noise admixture is not as fast. It is clear that the role of the product state supplement to the GHZ part is active and crucial. If an entangled state was tensored with the white noise, any local action on the additional particles would not bring any effect.  On the other hand, the supplement of the pure product state of $(N-3)$ qubits is optimal, as it allows to filter out the noise most efficiently.

\subsection{Bell inequalities and conjecture for arbitrary $N$}

We can recover analytically all the results presented in Tab.~\ref{product} including the particular states $\ket{\psi_3^N} = \ket{{\rm GHZ}_3}\ket{0...0}$, which are the most resistant to white noise for $N=4,5,6$. Starting from particular inequalities derived from the numerical method output for various $N$ and $r$, we deduce a general form of the inequality, which is given in the following iterative way
\begin{equation}
\langle C_N \rangle = \langle (1-A^{(N)}_1) C_{N-1} + L_r A^{(N)}_1 \rangle \leq L_r, \label{ineq_gen}
\end{equation}
where the iteration starts from $N-1=r$, where $C_r$ is a Bell expression with local bound $L_r$ involving the first $r$ particles. Note that in the above Bell inequality the last ($N-r$) observers perform only a single measurement $A^{(k)}_1~(k=r+1,\dots,N)$. The Bell expression (\ref{ineq_gen}) for $N\ge r$ can be explicitly written as
\begin{equation}
\label{closedform}
C_N = C_r + (L_r-C_r)f(N,r),
\end{equation}
where $f(N,r)$ is given as follows
\begin{eqnarray}
f(N,r) &=&\sum_{i=r+1}^N A^{(i)}_1 - \sum_{r+1 \leq i < j \leq N} A^{(i)}_1 A^{(j)}_1+
\sum_{r+1 \leq i < j < k \leq N} A^{(i)}_1 A^{(j)}_1 A^{(k)}_1 \\ &-& \cdots  + (-1)^{N+1} A^{(r+1)}_1 \ldots A^{(N)}_1.\nonumber
\end{eqnarray}

Let $C_r$ be the $r$-partite  Mermin-Ardehali-Belinskii-Klyshko (MABK) inequality \cite{Mermin,Ardehali,BK} recursively defined as
\begin{equation}
\label{MABKineq}
C_r = \frac{C_{r-1}}{2}\left(A_1^{(r)}+A_2^{(r)}\right)+\frac{C'_{r-1}}{2}\left(A_1^{(r)}-A_2^{(r)}\right),
\end{equation}
where the iteration starts with $C_1=A_1^{(1)}$ and $C'_r$ is derived from $C_r$ by swapping all the observables $A_1^{(i)}\leftrightarrow A_2^{(i)}$, $i=(1,\ldots,r)$ in the sum. The local realistic bound $L_r$ is equal to 1, whereas the maximum quantum value $Q_r$ is $2^{(r-1)/2}$ and can be achieved by the state $|GHZ_r\rangle$.

Let the observers $i=1,\ldots,r$ choose their measurements to maximize the MABK expression, and the rest as $A_1^{(i)} = -\sigma_z$ for $r<i\leq N$. We obtain $\avg{A_1^{(i)}}_{\psi_r^N} =-1$ and the expectation value $\avg{C_r}_{\psi_r^N} = 2^{(r-1)/2}$. By evaluating the $f(N,r)$ function for the above choice, we get $f(N,r)=2^{N-r}-1$. Hence,
we have $\avg{C_N} = \avg{C_r} + (L_r-\avg{C_r})(2^{N-r}-L_r)$ and the critical visibility $v_{crit} = \frac{L_r}{\avg{C_N}}$ becomes
\begin{equation}
v_{crit}^{\psi_r^N} = \frac{1}{2^N\left(2^{-\frac{r+1}{2}}-2^{-r}\right)+1}.
\label{vrn}
\end{equation}
This formula reproduces all the numerical values obtained in Tab.~\ref{product} with four-digit accuracy. Since the function $\left(2^{-\frac{r+1}{2}}-2^{-r}\right)$ takes its maximum for $r=3$, the lowest visibility is obtained for the states $\ket{\psi_3^N} = \ket{{\rm GHZ}_3}\ket{0...0}$ ($N > 3$).
Based on compatibility of Eq. (\ref{vrn}) with the numerical results presented in Tab.~\ref{product}, we conjecture that among any $N$-qubit states that state is the most robust against white noise admixture. Note that for $r=2$, $C_r$ reduces to the CHSH inequality and we obtain the inequality (\ref{ineq_dicke}) used in Sec. \ref{family}. We should stress that one can obtain the same result (\ref{vrn}) analyzing the violation of $C_r$ inequality by the $r$-qubit state resulting from the conditional measurements performed on the last ($N-r$) qubits \cite{SEN}.

Finally, let us invoke a recent method~\cite{Edepth} based on multipartite Bell inequalities, which detects entanglement depth~\cite{GTB} (a measure of genuine multipartite entanglement) in a device-independent way. Since our inequalities~(\ref{closedform}) (parametrized by $N$ and $r$) are maximally violated by an $r$-party GHZ state augmented with $(N-r)$ qubits in a product form, they are not capable to detect an entanglement depth bigger than $r$. 
Hence, our Bell tests can be considered as converse to the tests in Ref.~\cite{Edepth} which are rather suited to detect high entanglement depth, while the main benefit of our method is the detection of any entanglement in a very noisy environment. The critical visibility to violate the inequality presented in Ref. \cite{Edepth}, e. g.,  for the state $\ket{\psi_3^N}$ is equal to $3/5$ (independently on $N$), whereas the violation of our inequalities by the same state $\ket{\psi_3^N}$ can be observed in the presence of more noise. In our case, the critical visibility (\ref{vrn}) depends on $N$ and goes to 0 when $N$ tends to infinity.

\subsection{Other type of noise}

The phenomenon described above disappears when we consider a different noise admixture. For instance, for a product noise: $v \rho + (1-v)\rho_1 \otimes \cdots \otimes \rho_N$, where $\rho_i$ is the reduced density matrix of the $i$th subsystem of the state $\rho$, the critical visibilities obtained by means of the numerical method are equal to $2^{(N-r)/2} v_{crit}^{\rm GHZ}$ with the four-digit accuracy ($N=3,\dots,8$ and $r=2, \dots, N$).

\section{Conclusions}

We have analyzed quantum states, which offer the highest Bell inequality violation robustness against an admixture of the white noise. This robustness was studied in a specific experimental scenario, with two measurement settings per side.
We numerically prove up to $N=6$ and conjecture for an arbitrary $N$ that
the optimal states are products of three-qubit GHZ states and $(N-3)$ pure single qubit states. The corresponding inequalities represent a protocol, in which $(N-3)$ observers attempt to locally project the states, while the remaining three conduct a GHZ experiment. This result challenges common, but contradictory beliefs: that entanglement distributed among many parties is more strongly nonclassical, and that the robustness against white noise is a proper measure of quantumness.



\ack

W.L. is supported by NCN Grant No. 2014/14/M/ST2/00818. M.W. is supported by NCN Grant No. 2012/05/E/ST2/02352. T.V. acknowledges financial support from a J\'anos Bolyai Grant of the Hungarian Academy of Sciences,
the Hungarian National Research Fund OTKA (K111734), and SEFRI (COST action MP1006).


\appendix

\section{A single Bell inequality violated by any pure entangled state}

\label{app:ineq}

Our construction is based on the inequality~(\ref{ineq_gen}) with $r=2$ and $C_2$ given by the CHSH inequality. It is not difficult to see that this inequality is equivalent to the following one:
\begin{equation}
CH^{(1,2)}\prod_{k=3}^{N}p(A^{(k)}_1)\leq 0,
\label{Wineq}
\end{equation}
where $CH^{(i,j)} = p(A^{(i)}_2, A^{(j)}_2) + p(A^{(i)}_2, A^{(j)}_3) + p(A^{(i)}_3, A^{(j)}_2) - p(A^{(i)}_3, A^{(j)}_3) - p(A_2^{(i)}) - p(A_3^{(j)})$ is the Clauser-Horne expression \cite{CH}, $p(A_k^{(i)})$ denotes the probability of obtaining a ``1'' result by the $i$-th observer choosing observable $A_k$ and $p(A^{(i)}_l, A^{(j)}_m)$ is the probability of obtaining ``1'' by $i$th and $j$th observers choosing $A_l$ and $A_m$ observable, respectively.

From the above Bell inequality~(\ref{Wineq}), we construct the following symmetrized one (for any $N\ge 3$):
\begin{equation}
\sum_{1\le i<j\le N}{CH^{(i,j)}\prod_{k=1; k\neq i,j}^N p(A^{(k)}_1)}\leq 0.
\label{Wineqsym}
\end{equation}
This inequality involves three binary outcome settings ($A_1, A_2, A_3$) for each party. For the simplest case of three observables ($N=3$), it looks as follows:
\begin{equation}
CH^{(1,2)}p(A^{(3)}_1)+CH^{(1,3)}p(A^{(2)}_1)+CH^{(2,3)}p(A^{(1)}_1)\leq 0.
\label{Wineqsym3}
\end{equation}

In order to demonstrate that the single N-party Bell inequality~(\ref{Wineqsym}) is violated by any N-party pure entangled states, we make use of the result of Popescu and Rohrlich \cite{PR} (see also \cite{Cavalcanti}). They showed that for any N-party pure entangled state there exist $N-2$ local projections which leave the remaining two systems (say, systems $i$ and $j$) in a pure entangled state. Since any pure entangled states violate the Clauser-Horne (CH) inequality~\cite{Gisin}, it implies that inequality~(\ref{Wineq}) is violated by this state (where the successful projections are associated with outcomes ``1''). The fact that $i$ and $j$ can be any two systems out of the N systems is captured by the symmetrized inequality~(\ref{Wineqsym}). Indeed, let the parties $i$ and $j$ output 0 for their first measurement (i.e. they carry out a degenerate measurement resulting in $p(A^{(i)}_1)=0$ and $p(A^{(j)}_1)=0$), whereas the successful projections on the rest of the parties are associated with outcome ``1''. In this case the only term which survives in (\ref{Wineqsym}) is the one of (\ref{Wineq}), the other terms giving zero contribution.

Though there exists a more economical Bell inequality in terms of number of settings (it consists of only two settings per party) which is violated by any pure entangled states~\cite{Yu}, we believe the present proof based on our new inequality~(\ref{Wineqsym}) is simpler.

\end{document}